# DISPERSION OF ELECTROMAGNETIC WAVES IN A COAXIAL LINE FILLED WITH FERRITE

*Balakirev V.A., Onishchenko I.N.*

*National Science Center "Kharkov Institute of Physics and Technology", Kharkiv, Ukraine*
*E-mail: onish@kipt.kharkov.ua*

By solving Maxwell's equations the exact dispersion equation for electromagnetic waves propagating in a layered coaxial ferrite line is obtained. In particular the analytical consideration is carried out for a simpler case of complete filling of the coaxial line with ferrite (i.e. homogeneous ferrite line). The behavior of dispersion curves of $TEM$ - electromagnetic waves, as well as $E$ and $H$ - waveguide electromagnetic waves, is investigated.

## Introduction

When shock electromagnetic waves propagate in a ferrite coaxial transmission line [1-3], the effect of the microwave radiation generation by the leading edge of the shock wave occurs [4-9]. Gigawatt-power microwave generators based on ferrite lines differ from traditional microwave generators, for example, based on relativistic electron beams, in their simple design and compactness. Meanwhile, an electrodynamic theory of this phenomenon has not yet been developed. Only an important role of the phase synchronism $v_{sh} = v_{ph} \equiv \omega/k(\omega)$ between the front of the shock electromagnetic wave and the excited microwave wave is noted [1], where $v_{sh}$ is the velocity of the shock wave front, is the phase velocity of the excited electromagnetic wave, $\omega$ is the frequency, and $k(\omega)$ is the longitudinal wave number. The first step in developing a theory of microwave generators based on nonlinear ferrite lines is to study the electromagnetic dispersion properties of coaxial ferrite lines. This paper makes a step in this direction. A dispersion equation describing the dispersion characteristics of a coaxial ferrite line is obtained and studied. It should be noted that a wide range of issues on the linear theory of propagation of electromagnetic waves in ferrite media is covered, for example, in monographs [10-14].

**1. Statement of the problem. Obtaining the dispersion equation of electromagnetic waves**

The coaxial line is made in the form of two coaxial perfectly conducting cylindrical conductors. The radius of the inner conductor is $a$, and the outer one is $b$. The region $r_f > r > a$ adjacent to the inner conductor is filled with ferrite with a magnetic permeability tensor $\hat{\mu}$ and a permittivity $\varepsilon_f$. The space between the ferrite and the outer conductor $b > r > r_f$ is filled with a dielectric with a permittivity $\varepsilon_d$. The high-frequency magnetic properties of the ferrite are described by the magnetic permeability tensor

$$\hat{\mu} = \begin{pmatrix} \mu_\perp & -i\mu_a & 0 \\ i\mu_a & \mu_\perp & 0 \\ 0 & 0 & 1 \end{pmatrix}, \qquad (1)$$

where $\mu_\perp = 1 + (\mu_0 - 1)\dfrac{\omega_h^2}{\omega_h^2 - \omega^2} \equiv \dfrac{\omega^2 - \mu_0\omega_h^2}{\omega^2 - \omega_h^2}$, $\mu_a = (\mu_0 - 1)\dfrac{\omega\omega_h}{\omega_h^2 - \omega^2}$, $\omega_h = eH_0/mc$ is the precession frequency of the ferrite magnetization vector, $\mu_0 = 1 + 4\pi\chi_0$ is static magnetic permeability, $\chi_0 = M_s/H_0$ is magnetic susceptibility, $M_s$ is saturation magnetization, and $H_0$ is external magnetic field strength. The propagation of monochromatic traveling electromagnetic waves in a ferrite coaxial line is described by the system of Maxwell equations

$$rot\vec{E} = ik_0\vec{B}, \quad rot\vec{H} = -ik_0\vec{D}, \tag{2}$$

$$div\vec{D} = 0, \quad div\vec{B} = 0. \tag{3}$$

where $\vec{D} = \varepsilon_f \vec{E}$ is the electric induction, $\varepsilon_f$ the permittivity of ferrite, the magnetic induction vector, . Let us write out the system of Maxwell equations (2), (3) taking into account the expression for the magnetic permeability tensor (1) in components

$$\frac{1}{r}\frac{d}{dr}rH_\varphi = -ik_0\varepsilon_f E_z, \quad kH_\varphi = k_0\varepsilon_f E_r, \quad \frac{dE_z}{dr} - ikE_r = -ik_0\left(i\mu_a H_r + \mu_\perp H_\varphi\right), \tag{4}$$

$$\frac{1}{r}\frac{d}{dr}rE_\varphi = ik_0 H_z, \quad kE_\varphi = -k_0\left(\mu_\perp H_r - i\mu_a H_\varphi\right), \quad \frac{dH_z}{dr} - ikH_r = ik_0\varepsilon_f E_\varphi, \tag{5}$$

$$\frac{1}{r}\frac{d}{dr}rE_r + ikE_z = 0, \quad \mu_\perp\frac{1}{r}\frac{d}{dr}rH_r - i\mu_a\frac{1}{r}\frac{d}{dr}rH_\varphi + ikH_z = 0, \tag{6}$$

$k$ is longitudinal wave number. In turn, the system of equations (4) - (6) can be reduced to two coupled ordinary differential equations for the longitudinal components of the electromagnetic field

$$\Delta_\perp E_z + \kappa_e^2 E_z = -igH_z, \quad \Delta_\perp H_z + \kappa_h^2 H_z = i\varepsilon_f g E_z \tag{7}$$

or to one fourth-order equation, for example, for the longitudinal component of the electric field

$$\Delta_\perp^2 E_z + \left(\kappa_e^2 + \kappa_h^2\right)\Delta_\perp E_z + \left(\kappa_e^2\kappa_h^2 - \varepsilon_f g^2\right)E_z = 0. \tag{8}$$

Here $\Delta_\perp = \frac{1}{r}\frac{d}{dr}r\frac{d}{dr}$, $\kappa_e^2 = k_0^2\varepsilon_f\mu_{eff} - k^2$, $\mu_{eff} = \frac{\mu_\perp^2 - \mu_a^2}{\mu_\perp} \equiv \frac{\omega^2 - \mu_0^2\omega_h^2}{\omega^2 - \mu_0\omega_h^2}$, $\kappa_h^2 = k_0^2\varepsilon_f - \frac{k^2}{\mu_\perp}$,

$$g = k_0 k \frac{\mu_a}{\mu_\perp} = (\mu_0 - 1)k_0 k \frac{\omega\omega_h}{\mu_0\omega_h^2 - \omega^2}.$$

In the dielectric region $b > r > r_f$, the system of coupled equations (7) splits into two independent equations describing the $E$-waves

$$\Delta_\perp E_z + \lambda_d^2 E_z = 0 \tag{9a}$$

and $H$-waves

$$\Delta_\perp H_z + \lambda_d^2 H_z = 0, \tag{9b}$$

where $\lambda_d^2 = k_0^2\varepsilon_d - k^2$. The system of equations (7) must be supplemented with boundary conditions. The tangential components of the electric field vanish on the walls of a perfectly conducting coaxial line

$$E_{fz,\varphi}(r = a) = 0, \quad E_{dz,\varphi}(r = b) = 0, \tag{10}$$

At the interface between the media $r = r_f$, the tangential components of the electric and magnetic fields are continuous

$$E_{fz,\varphi}(r = r_f) = E_{dz,\varphi}(r = r_f), \quad H_{fz,\varphi}(r = r_f) = H_{dz,\varphi}(r = r_f). \tag{11}$$

The subscript $f$ in the field components included in the boundary conditions (10), (11) means that the fields belong to the ferrite, and the index $d$ means that they belong to the dielectric. The solution to equation (8) is a linear superposition of cylindrical functions, which is convenient to represent in the form

$$E_{fz} = A_{e1}\eta_0(\lambda_1 r) + A_{e2}\varsigma_0(\lambda_1 r) + A_{h1}\eta_0(\lambda_2 r) + A_{h2}\varsigma_0(\lambda_2 r),$$

$$H_{fz} = \frac{i}{g}\left\{\Gamma_e\left[A_{e1}\eta_0(\lambda_1 r) + A_{e2}\varsigma_0(\lambda_1 r)\right] + \Gamma_h\left[A_{h1}\eta_0(\lambda_2 r) + A_{h2}\varsigma_0(\lambda_2 r)\right]\right\},$$

where

$$\eta_0(\lambda r) \equiv \frac{\Delta_{00}(\lambda r, \lambda a)}{\Delta_{00}(\lambda r_f, \lambda a)}, \quad \varsigma_0(\lambda r) \equiv \frac{J_0(\lambda r)}{J_0(\lambda a)},$$

$$\Delta_{mn}(\lambda r, \lambda a) = J_m(\lambda r) N_n(\lambda a) - J_n(\lambda a) N_m(\lambda r), \quad \Gamma_{1,2} = \kappa_e^2 - \lambda_{1,2}^2.$$

The transverse wave numbers are the roots of the biquadratic equation

$$\lambda^4 - \left(\kappa_e^2 + \kappa_h^2\right)\lambda^2 + \kappa_e^2 \kappa_h^2 - \varepsilon_f g^2 = 0. \tag{12}$$

The roots of this equation have form

$$\lambda_{1,2}^2 = \frac{1}{2}\left[\kappa_e^2 + \kappa_h^2 \pm \sqrt{\left(\kappa_e^2 - \kappa_h^2\right)^2 + 4\varepsilon_f g^2}\right], \tag{13}$$

$$\kappa_e^2 - \kappa_h^2 \equiv \frac{\omega_h^2(\mu_0 - 1)}{\mu_0 \omega_h^2 - \omega^2}\kappa_0^2, \quad \kappa_e^2 + \kappa_h^2 \equiv \frac{2\omega^2 \kappa^2 - (\mu_0 + 1)\omega_h^2 \kappa_0^2}{\omega^2 - \mu_0 \omega_h^2},$$

where $\kappa_0^2 = k_0^2 \varepsilon_f \mu_0 - k^2$, $\kappa^2 = k_0^2 \varepsilon_f - k^2$. Expression (13) is equivalent to the following

$$\lambda_{1,2}^2 = \frac{1}{2(\mu_0 \omega_h^2 - \omega^2)}\left[(\mu_0 + 1)\omega_h^2 \kappa_0^2 - 2\omega^2 \kappa^2 \pm (\mu_0 - 1)\omega_h \sqrt{\omega_h^2 \kappa_0^4 + 4\omega^2 k_0^2 \varepsilon_f k^2}\right],$$

Note that in formula (13) the expression under the radical is always positive. Therefore, unlike a magnetoactive plasma waveguide, a ferrite waveguide does not have complex (volume-surface) eigenwaves, which are inherent in magnetoactive plasma waveguides [13, 14].

The expressions for the azimuthal components of the electromagnetic field follow from the first equations of system (4), (5) and have the form

$$H_{f\varphi} = -ik_0 \varepsilon_f \left\{\frac{1}{\lambda_1^2}\left[A_{e1}\eta_1(\lambda_1 r) + A_{e2}\varsigma_1(\lambda_1 r)\right] + \frac{1}{\lambda_2^2}\left[A_{h1}\eta_1(\lambda_2 r) + A_{h2}\varsigma_1(\lambda_2 r)\right]\right\},$$

$$E_{f\varphi} = -\frac{k_0}{g}\left\{\frac{\Gamma_1}{\lambda_1^2}\left[A_{e1}\eta_1(\lambda_1 r) + A_{e2}\varsigma_1(\lambda_1 r)\right] + \frac{\Gamma_2}{\lambda_2^2}\left[A_{h1}\eta_1(\lambda_2 r) + A_{h2}\varsigma_1(\lambda_2 r)\right]\right\}.$$

Here

$$E_{dz} = B_e \frac{\Delta_{00}(\lambda_d r, \lambda_d b)}{\Delta_{00}(\lambda_d r_f, \lambda_d b)}, \quad H_{dz} = B_h \frac{\Delta_{01}(\lambda_d r, \lambda_d b)}{\Delta_{01}(\lambda_d r_f, \lambda_d b)},$$

$$H_{d\varphi} = -i\frac{k_0 \varepsilon_d}{\lambda_d} B_e \frac{\Delta_{10}(\lambda_d r, \lambda_d b)}{\Delta_{00}(\lambda_d r_f, \lambda_d b)}, \quad E_{d\varphi} = i\frac{k_0}{\lambda_d} B_h \frac{\Delta_{11}(\lambda_d r, \lambda_d b)}{\Delta_{01}(\lambda_d r_f, \lambda_d b)}.$$

In the dielectric region, the solutions of equations (9) satisfying the boundary conditions (10) on the outer wall of the coaxial line have the form

$$E_{dz} = B_e \frac{\Delta_{00}(\lambda_d r, \lambda_d b)}{\Delta_{00}(\lambda_d r_f, \lambda_d b)}, \quad H_{dz} = B_h \frac{\Delta_{01}(\lambda_d r, \lambda_d b)}{\Delta_{01}(\lambda_d r_f, \lambda_d b)},$$

$$H_{d\varphi} = -i\frac{k_0 \varepsilon_d}{\lambda_d} B_e \frac{\Delta_{10}(\lambda_d r, \lambda_d b)}{\Delta_{00}(\lambda_d r_f, \lambda_d b)}, \quad E_{d\varphi} = i\frac{k_0}{\lambda_d} B_h \frac{\Delta_{11}(\lambda_d r, \lambda_d b)}{\Delta_{01}(\lambda_d r_f, \lambda_d b)}.$$

To obtain the dispersion equation, we use the boundary conditions (11) at the interface between the media, as well as the boundary conditions (10) on the inner conductor. As a result, we obtain the following system of linear equations for determining the coefficients $A_{e1,2}, A_{h1,2}$

$$A_{e2} + A_{h2} = 0, \tag{14a}$$

$$\Gamma_1 \lambda_2^2 \left[A_{e1}\eta_1(\lambda_1 a) + A_{e2}\varsigma_1(\lambda_1 a)\right] + \Gamma_2 \lambda_1^2 \left[A_{h1}\eta_1(\lambda_2 a) + A_{h2}\varsigma_1(\lambda_2 a)\right] = 0, \tag{14b}$$

$$\frac{\Lambda}{\varepsilon_f} \frac{A_{e1}\eta_0(\lambda_1 r_f) + A_{e2}\varsigma_0(\lambda_1 r_f) + A_{h1}\eta_0(\lambda_2 r_f) + A_{h2}\varsigma_0(\lambda_2 r_f)}{\lambda_2^2\left[A_{e1}\eta_1(\lambda_1 r_f) + A_{e2}\varsigma_1(\lambda_1 r_f)\right] + \lambda_1^2\left[A_{h1}\eta_1(\lambda_2 r_f) + A_{h2}\varsigma_1(\lambda_2 r_f)\right]} = Z_E, \tag{14c}$$

$$\frac{\Gamma_1\lambda_2^2\left[A_{e1}\eta_1(\lambda_1 r_f)+A_{e2}\varsigma_1(\lambda_1 r_f)\right]+\Gamma_2\lambda_1^2\left[A_{h1}\eta_1(\lambda_2 r_f)+A_{h2}\varsigma_1(\lambda_2 r_f)\right]}{\Gamma_1\left[A_{e1}\eta_0(\lambda_1 r_f)+A_{e2}\varsigma_0(\lambda_1 r_f)\right]+\Gamma_2\left[A_{h1}\eta_0(\lambda_2 r_f)+A_{h2}\varsigma_0(\lambda_2 r_f)\right]}=\Lambda Z_H, \quad (14d)$$

where

$$Z_E=\frac{\lambda_d}{\varepsilon_d}\frac{\Delta_{00}(\lambda_d r_f,\lambda_d b)}{\Delta_{10}(\lambda_d r_f,\lambda_d b)},\quad Z_H=\frac{1}{\lambda_d}\frac{\Delta_{11}(\lambda_d r_f,\lambda_d b)}{\Delta_{01}(\lambda_d r_f,\lambda_d b)},\quad \Lambda=\lambda_1^2\lambda_2^2=\kappa_e^2\kappa_h^2-\varepsilon_f g^2.$$

The parameters $Z_E$ and $Z_H$ have a simple physical meaning. The quantity $\frac{i}{k_0}Z_E$ is the impedance for the $E$ - wave, and $ik_0 Z_H$ is the impedance for the $H$ - wave at the interface. Thus, we have obtained a system of four linear equations, from the solvability condition of which we obtain a dispersion equation describing the dispersion properties of a layered coaxial ferrite line

$$\begin{vmatrix} 0 & 1 & 0 & 1 \\ \Gamma_1\lambda_2^2\eta_1(\lambda_1 a) & \Gamma_1\lambda_2^2\varsigma_1(\lambda_1 a) & \Gamma_2\lambda_1^2\eta_1(\lambda_2 a) & \Gamma_2\lambda_1^2\varsigma_1(\lambda_2 a) \\ D_{\eta e} & D_{\varsigma e} & D_{\eta h} & D_{\varsigma h} \\ \Gamma_1 B_{\eta e} & \Gamma_1 B_{\varsigma e} & \Gamma_2 B_{\eta h} & \Gamma_2 B_{\varsigma h} \end{vmatrix}=0. \quad (15)$$

Here

$$D_{\eta 0}=1-Q_1,\quad D_{\eta 1}=\lambda_1^2\eta_1(\lambda_2 r_f)-\lambda_2^2 Q_1\eta_1(\lambda_1 r_f),$$
$$D_{\varsigma 0}=\varsigma_0(\lambda_2 r_f)-\varsigma_0(\lambda_1 r_f)-Q_2,$$
$$D_{\varsigma 1}=\lambda_1^2\varsigma_1(\lambda_2 r_f)-\lambda_2^2\varsigma_1(\lambda_1 r_f)-\lambda_2^2 Q_2\eta_1(\lambda_1 r_f),$$
$$B_{\eta 0}=\Gamma_h-\Gamma_e Q_1,\quad B_{\eta 1}=\Gamma_2\lambda_1^2\eta_1(\lambda_2 r_f)-\Gamma_1\lambda_2^2 Q_1\eta_1(\lambda_1 r_f),$$
$$B_{\varsigma 0}=\Gamma_2\varsigma_0(\lambda_2 r_f)-\Gamma_1\left[\varsigma_0(\lambda_1 r_f)+Q_2\right],$$
$$B_{\varsigma 1}=\Gamma_2\lambda_1^2\varsigma_1(\lambda_2 r_f)-\Gamma_1\lambda_2^2\left[\varsigma_1(\lambda_1 r_f)+Q_2\eta_1(\lambda_1 r_f)\right],$$
$$Q_1=\frac{\Gamma_2\lambda_1^2}{\Gamma_1\lambda_2^2}\frac{\eta_1(\lambda_2 a)}{\eta_1(\lambda_1 a)},\quad Q_2=Q_1\xi_2-\xi_1,\quad \xi_{1,2}=\frac{\varsigma_1(\lambda_{1,2}a)}{\eta_1(\lambda_{1,2}a)}.$$

## 2. Coaxial line completely filled with ferrite

The dispersion equation in the form (15) is cumbersome and does not lend itself to a complete analytical consideration. Therefore, below we study it in limiting cases. The ferrite coaxial line of the simplest geometry is a line completely filled with ferrite $r_f=b$. Accordingly, the dielectric layer is absent. For such a line, the impedances are $Z_{E,H}=0$. In this case, the general dispersion equation (15) splits into two independent dispersion equations

$$\Lambda(\omega,k)\Delta_{wg}(\omega,k)=0,$$

where

$$\Delta_{wg}(\omega,k)=\begin{vmatrix} 0 & 1 & 0 & 1 \\ \Gamma_1\lambda_2^2\eta_1(\lambda_1 a) & \Gamma_1\lambda_2^2\varsigma_1(\lambda_1 a) & \Gamma_2\lambda_1^2\eta_1(\lambda_2 a) & \Gamma_2\lambda_1^2\varsigma_1(\lambda_2 a) \\ 1 & \varsigma_0(\lambda_1 b) & 1 & \varsigma_0(\lambda_2 b) \\ \Gamma_1\lambda_2^2\eta_1(\lambda_1 b) & \Gamma_1\lambda_2^2\varsigma_1(\lambda_1 b) & \Gamma_2\lambda_1^2\eta_1(\lambda_2 b) & \Gamma_2\lambda_1^2\varsigma_1(\lambda_2 b) \end{vmatrix}. \quad (17)$$

The first of them

$$\Lambda(\omega,k) \equiv \kappa_e^2 \kappa_h^2 - \varepsilon_0 g^2 = 0 \tag{18}$$

describes the dispersion of - waves, and the second equation

$$\Delta_{wg}(\omega,k) = 0 \tag{19}$$

describes the dispersion of the waveguide electromagnetic eigen waves of a coaxial ferrite waveguide.

## 2.1. Dispersion properties of *TEM* - wave

Let us dwell in more detail on the dispersion properties of the *TEM* - waves of the coaxial line under consideration. The dispersion equation for -waves (18) can be transformed to the form

$$k_0^2 \varepsilon_0 (\mu_\perp + \mu_a) = k^2.$$

This equation is equivalent to the following

$$\frac{\omega^2}{c^2} \varepsilon_f \frac{\omega - \mu_0 \omega_h}{\omega - \omega_h} = k^2. \tag{20}$$

Directly from the dispersion equation (20) follow the dependences of the longitudinal wave number and phase velocity of the *TEM* - wave on the frequency

$$k = \frac{\omega}{c} \sqrt{\varepsilon_f} \sqrt{\frac{\mu_0 \omega_h - \omega}{\omega_h - \omega}},$$

$$v_{ph} \equiv \frac{\omega}{k} = \frac{c}{\sqrt{\varepsilon_f}} \sqrt{\frac{\omega_h - \omega}{\mu_0 \omega_h - \omega}}. \tag{21}$$

It is convenient to introduce the dimensionless frequency and longitudinal wave number

$$w = \omega / \omega_h, \quad K = kc / \omega_h \sqrt{\varepsilon_f}. \tag{22}$$

The dispersion equation (20) in these dimensionless variables takes the form

$$w^2 \frac{\mu_0 - w}{1 - w} = K^2.$$

The dispersion dependences in the selected dimensionless variables depend only on one parameter - the static magnetic permeability $\mu_0$. The cutoff frequency $\omega_{cut} = \mu_0 \omega_h$. follows directly from the dispersion equation (20). *TEM* - waves exist in the frequency ranges $\omega < \omega_h$ and $\omega > \omega_h \mu_0$. In the opacity frequency band, *TEM* - waves are absent. In the long-wave approximation $K \ll 1$, we obtain two solutions

$$\omega = k v_F - k^2 v_F^2 \frac{\mu_0 - 1}{2 \omega_h \mu_0}, \tag{23}$$

$$\omega = \mu_0 \omega_h + k^2 \frac{c^2}{\omega_h \varepsilon_f} \frac{\mu_0 - 1}{\mu_0^2}, \tag{24}$$

where $v_F = c / \sqrt{\varepsilon_f \mu_0}$. With increasing wave number, the linear growth of the frequency of the low-frequency branch (23) of the coaxial wave slows down, and the phase velocity of the *TEM* - wave decrease

$$v_{ph} = \frac{\omega}{k} = \frac{c}{\sqrt{\varepsilon_0 \mu_0}} - \frac{\mu_0 - 1}{2 \mu_0^2 \varepsilon_0} \frac{c^2}{\omega_h} k.$$

At $k \to \infty$ the frequency of the LF frequency *TEM* - wave asymptotically approaches the precession frequency $\omega_h$ from below according to the law

$$\omega = \omega_h - \frac{\omega_h^3}{k^2 c^2} \varepsilon_f (\mu_0 - 1),$$

and its phase velocity (21) tends to zero. As for the high-frequency branch of the coaxial wave, in accordance with formula (24), it begins with the HF cutoff frequency. With the growth of the longitudinal wave number, its frequency increases, and at $K \to \infty$ the dispersion curve asymptotically goes to a straight line

$$\omega = \frac{kc}{\sqrt{\varepsilon_f}} + \frac{\mu_0 - 1}{2}\omega_h$$

in law

$$\omega = \frac{kc}{\sqrt{\varepsilon_f}} + \frac{\mu_0 - 1}{2}\omega_h - \frac{\mu_0}{2}\left(1 + \frac{\mu_0}{4}\right)\frac{\omega_h \sqrt{\varepsilon_f}}{c}\frac{1}{k}.$$

Thus, the dispersion curve of the *TEM* - wave in a completely filled ferrite line has a low-frequency and a high-frequency branch. These branches of the *TEM* - waves are separated by an opacity band.

### 2.2. Dispersion of waveguide electromagnetic waves

Let us proceed to the study of the dispersion properties of waveguide electromagnetic waves of a ferrite coaxial line. The dispersion equation for these waves has the form (19). The determinant (17) can be expanded and the dispersion equation (19) can be presented in the form

$$\Gamma_2^2 \lambda_1^2 \Delta_{00}(\lambda_1 b, \lambda_1 a)\Delta_{11}(\lambda_2 b, \lambda_2 a) + \Gamma_1^2 \lambda_2^2 \Delta_{00}(\lambda_2 b, \lambda_2 a)\Delta_{11}(\lambda_1 b, \lambda_1 a) =$$

$$= \Gamma_1\Gamma_2\lambda_1\lambda_2\left[\frac{8}{\pi^2 \lambda_1\lambda_2 ab} + \Delta_{10}(\lambda_1 b, \lambda_1 a)\Delta_{01}(\lambda_2 b, \lambda_2 a) + \Delta_{10}(\lambda_2 b, \lambda_2 a)\Delta_{01}(\lambda_1 b, \lambda_1 a)\right]. \quad (25)$$

Let's introduce the gyrotropy parameter

$$\rho_f = \frac{\varepsilon_f g^2}{\left(\kappa_e^2 - \kappa_h^2\right)^2} \equiv \varepsilon_f \frac{\omega^2}{\omega_h^2}\frac{k_0^2 k^2}{\kappa_0^4} \quad (26)$$

and we represent expression (13) for transverse wave numbers in the form

$$\lambda_{1,2}^2 = \frac{1}{2}\left[\kappa_e^2 + \kappa_h^2 \pm \left(\kappa_e^2 - \kappa_h^2\right)\sqrt{1 + 4\rho_f}\right]. \quad (27)$$

Let us consider the case of weak gyrotropy of ferrite, when the parameter is small

$$4\rho_f \ll 1. \quad (28)$$

In this approximation, the expressions for the transverse wave numbers (27) take the form

$$\lambda_{1,2}^2 = \kappa_{e,h}^2 \pm \delta\lambda^2, \quad (29)$$

where

$$\delta\lambda^2 = \rho_f\left(\kappa_e^2 - \kappa_h^2\right) \equiv \varepsilon_f\left(\mu_0 - 1\right)\frac{\omega^2}{\mu_0\omega_h^2 - \omega}\frac{k_0^2 k^2}{\kappa_0^2}. \quad (30)$$

Accordingly, we have the following relationships $\Gamma_1 = -\delta\lambda^2$, $\Gamma_2 = \kappa_e^2 - \kappa_h^2$, $\frac{\Gamma_1}{\Gamma_2} = -\rho_f$. In this approximation, the dispersion equation (25) can be simplified and represented as

$$\Delta_{00}(\lambda_1 b, \lambda_1 a)\Delta_{11}(\lambda_2 b, \lambda_2 a) =$$

$$= -\rho_f\frac{1}{\lambda_1^2}\left[\frac{8}{\pi^2 ab} + \lambda_1\lambda_2\left[\Delta_{10}(\lambda_1 b, \lambda_1 a)\Delta_{01}(\lambda_2 b, \lambda_2 a) + \Delta_{10}(\lambda_2 b, \lambda_2 a)\Delta_{01}(\lambda_1 b, \lambda_1 a)\right]\right]. \quad (31)$$

The dispersion equation (31) describes weakly coupled *E* and *H* - waves of a coaxial ferrite waveguide. The linear bond between these types of waves is realized through the weak gyrotropy of ferrite. Taking into account condition (28), the dispersion equation (31) can be solved by the method of successive approximations in the parameter $\rho_f$. In the main approximation $\rho_f = 0$ (33), the dispersion equation (31) is decomposed into two independent equations

$$\Delta_{00}(\lambda_1 b, \lambda_1 a) \equiv J_0(\lambda_1 b)N_0(\lambda_1 a) - J_0(\lambda_1 a)N_0(\lambda_1 b) = 0, \qquad (32)$$

$$\Delta_{11}(\lambda_2 b, \lambda_2 a) \equiv J_1(\lambda_2 b)N_1(\lambda_2 a) - J_1(\lambda_2 a)N_1(\lambda_2 b) = 0. \qquad (33)$$

Equation (32) is the dispersion equation for $E$ - waves of a ferrite coaxial line in the long-wave region. Accordingly, equation (33) describes the dispersion of $H$ - waves. In the next approximation in parameter $\rho_f$, instead of equation (31), we obtain the following dispersion equations for $E$ - waves

$$\Delta_{00}(v_e \eta, v_e) = -\frac{8}{\pi^2} \rho_f \frac{Q_e(v_e, v_h)}{v_e^2 \eta}, \qquad (34)$$

$$Q_e(v_e, v_h) = \frac{1}{\Delta_{11}(v_h \eta, v_h)} \left\{ 1 + \frac{\pi}{4} v_h \left[ \frac{J_0(v_e)}{J_0(v_e \eta)} \Delta_{01}(v_h \eta, v_h) - \eta \frac{J_0(v_e \eta)}{J_0(v_e)} \Delta_{10}(v_h \eta, v_h) \right] \right\},$$

$v_{e,h} = \lambda_{1,2} a, \ \eta = b/a$

and for $H$ - waves

$$\Delta_{11}(v_h \eta, v_h) = -\frac{8}{\pi^2} \rho_f \frac{Q_h(v_h, v_e)}{v_e^2 \eta}, \qquad (35)$$

$$Q_h(v_h, v_e) = \frac{1}{\Delta_{00}(v_e \eta, v_e)} \left\{ 1 - \frac{\pi}{4} v_e \left[ \frac{J_0(v_h)}{J_0(v_h \eta)} \Delta_{01}(v_e \eta, v_e) - \eta \frac{J_0(v_h \eta)}{J_0(v_h)} \Delta_{10}(v_e \eta, v_e) \right] \right\}.$$

It follows from these equations that taking into account weak gyrotropy (small values of the parameter $\rho_f$) leads only to a shift in the transverse wave numbers. This means that in the approximation under consideration, the separation into $E$ and $H$ - waves is preserved.

Let us determine the cutoff frequencies and the behavior of the dispersion curves of electromagnetic waves in the vicinity of these frequencies. For, $k = 0$ we have $\rho_f = 0$ identically. The cutoff frequencies of $E$ - waves are determined by equation (32). From this equation follows an algebraic equation for determining the cutoff frequencies

$$k_0^2 \varepsilon_f \frac{\mu_\perp^2 - \mu_a^2}{\mu_\perp} = \frac{v_{en}^2}{a^2}. \qquad (36)$$

$v_{en}$ are roots of the transcendental equation

$$J_0(v_{en} \eta)N_0(v_{en}) - J_0(v_{en})N_0(v_{en} \eta) = 0. \qquad (37)$$

Under $v_{en} \gg 1$ for these roots we have the following approximate representation $v_{en} = \pi n/(\eta - 1)$. In dimensionless variables (22), equation (36) is equivalent to the following

$$w^2 \frac{w^2 - \mu_0^2}{w^2 - \mu_0} = w_{en}^2, \qquad (38)$$

where $w_{en} = \frac{\omega_{en}}{\omega_h}$, $\omega_{en} = \frac{v_{en} c}{a \sqrt{\varepsilon_f}}$. The biquadratic equation (38) has roots

$$w_{cen}^{(hf,lf)2} = \frac{1}{2} \left[ \mu_0^2 + w_{en}^2 \pm \sqrt{\left( \mu_0^2 + w_{en}^2 \right)^2 - 4\mu_0 w_{en}^2} \right]. \qquad (39)$$

Note that the expression under the radical in (39) is always positive. Thus, each radial harmonic of the $E$ - wave has a high-frequency (+ sign) and a low-frequency (sign ) cutoff frequency. This fact unambiguously indicates that the dispersion curve of each $E$ - radial harmonic has a low-frequency and a high-frequency branch. Each of these branches begins with the corresponding cutoff frequency. In the most interesting limiting case

$$w_{en}^2 \ll \mu_0 \qquad (40)$$

expressions (39) for the HF and LF - cutoff frequencies are simplified

$$w_{cen}^{(hf)2} = \mu_0^2 + \frac{\mu_0 - 1}{\mu_0} w_{en}^2, \tag{41}$$

$$w_{cen}^{(lf)} = \frac{w_{en}}{\sqrt{\mu_0}}. \tag{42}$$

In dimensional units, formulas (41), (42) take the form

$$\omega_{cen}^{(hf)2} = \omega_h^2 \mu_0^2 + \frac{\mu_0 - 1}{\mu_0} \omega_{en}^2, \tag{43}$$

$$\omega_{cen}^{(lf)} = \omega_{en} / \sqrt{\mu_0}. \tag{44}$$

Let us now consider the dispersion of $E$ - radial harmonics in the vicinity of their cutoff frequencies. Solving the dispersion equation for $E$ - electromagnetic waves (34) by the method of successive approximations, we find an approximate value of the transverse wave number

$$\lambda_1(\omega, k) = \frac{v_{en} + \Delta v_{en}(\omega, k)}{a}, \quad \Delta v_{en} = \frac{4}{\pi} \rho_f \frac{Q_{en}(v_h)}{\eta v_{en} j_{en}^{(0)}}, \tag{45}$$

$$Q_{en}(v_h) \equiv Q_e(v_{en}, v_h), \quad j_{en}^{(0)} = \frac{J_0(v_{en})}{J_0(v_{en}\eta)} - \frac{J_0(v_{en}\eta)}{J_0(v_{en})}.$$

Taking into account this approximate expression for the transverse wave number, we obtain the dispersion equation for the $E$ - radial harmonics, which is valid for the case of weak gyrotropy $\rho_f \ll 1$

$$\kappa_e^2 + \rho_f \left[ \kappa_e^2 - \kappa_h^2 - \frac{8}{\pi a^2 \eta} \frac{Q_{en}(v_h)}{j_{en}^{(0)}} \right] = \frac{v_{en}^2}{a^2}. \tag{46}$$

In the limiting case $k_h a \gg 1$, $k_h = \omega_h / c$, the last term in the right-hand side of this equation can be neglected, taking into account the shift of the eigenvalues $\Delta v_{en}$ (43). As a result, we obtain the following dispersion equation

$$\omega^2 \frac{\omega_h^2 \mu_0^2 - \omega^2}{\omega_h^2 \mu_0 - \omega^2} - \frac{k^2 c^2}{\varepsilon_f} + \frac{k^2}{k_0^2 \varepsilon_f \mu_0 - k^2} \frac{\omega^4(\mu_0 - 1)}{\omega_h^2 \mu_0 - \omega^2} = \omega_{en}^2. \tag{47}$$

Near the cutoff frequencies in the dispersion equation (47) it is sufficient to retain only the terms proportional to $k^2$. The dispersion equation in this approximation takes the form

$$\omega^2 \frac{\omega_h^2 \mu_0^2 - \omega^2}{\omega_h^2 \mu_0 - \omega^2} - \frac{k^2 c^2}{\varepsilon_f} \left( 1 - \frac{1}{\mu_0} \frac{\omega^2(\mu_0 - 1)}{\omega_h^2 \mu_0 - \omega^2} \right) = \omega_{en}^2. \tag{48}$$

In the limiting case (40), simple dispersion laws for the low-frequency and high-frequency $E$ - radial harmonics follow from this dispersion equation

$$\omega_{lf}^2 = \frac{\omega_{en}^2}{\mu_0} + \frac{k^2 c^2}{\varepsilon_f \mu_0} \left( 1 - \frac{\mu_0 - 1}{\mu_0^3} \frac{\omega_{en}^2}{\omega_h^2} \right), \tag{49}$$

$$\omega_{hf}^2 = \omega_h^2 + \frac{\mu_0 - 1}{\mu_0} \omega_{en}^2 + 2(\mu_0 - 1) \frac{k^2 c^2}{\mu_0 \varepsilon_f}. \tag{50}$$

Let us now consider the LF cutoff frequencies of $H$ - the waves and the dispersion laws in the vicinity of these frequencies. Solving the dispersion equation for $H$ - electromagnetic waves (35) by the method of successive approximations, we find the approximate value of the transverse wave number

$$\lambda_2(\omega, k) = \frac{v_{hn} + \Delta v_{hn}(\omega, k)}{a}, \quad \Delta v_{hn} = \frac{4}{\pi} \rho_f \frac{Q_{hn}(v_e)}{\eta v_e^2 j_{hn}^{(1)}} v_{hn}, \tag{51}$$

$$Q_{hn}(v_e) \equiv Q_h(v_e, v_{hn}), \quad j_{hn}^{(1)} = \frac{J_1(v_{hn})}{J_1(v_{hn}\eta)} - \frac{J_1(v_{hn}\eta)}{J_1(v_{hn})},$$

where $v_{hn}$ - are the roots of the transcendental equation

$$J_1(v_{hn}\eta)N_1(v_{hn}) - J_1(v_{hn})N_1(v_{en}\eta) = 0. \tag{52}$$

The approximate values of these roots have the same form as in the case of $E$ - waves $v_{hn} = \pi n/(\eta - 1)$. Taking into account the expression for the transverse wave number (43) and the values $\lambda_2$ (29), (30), we obtain the dispersion equation for $H$ - radial harmonics in the weakly gyrotropic case $\rho_f \ll 1$

$$\kappa_h^2 - \rho_f \left[\kappa_e^2 - \kappa_h^2 + \frac{8}{\pi a^2 \eta} \frac{v_{hn}}{v_e^2} \frac{Q_{hn}(v_e)}{j_{hn}^{(1)}}\right] = \frac{v_{hn}^2}{a^2}. \tag{53}$$

In the limiting case $k_h a \gg 1$, we can neglect the shift of the eigen numberss $\Delta v_{hn}$ (51) and represent this dispersion equation in the form

$$\omega^2 - \frac{k^2 c^2}{\varepsilon_f} \frac{\omega_h^2 - \omega^2}{\omega_h^2 \mu_0 - \omega^2} - \frac{k^2}{k_0^2 \varepsilon_f \mu_0 - k^2} \frac{\omega^4(\mu_0 - 1)}{\omega_h^2 \mu_0 - \omega^2} = \omega_{hn}^2, \tag{54}$$

$\omega_{hn} = \frac{cv_{hn}}{a\sqrt{\varepsilon_f}}$. From this equation it follows that each $H$ - radial harmonic has only one cutoff frequency

$$\omega_{chn} = \omega_{hn} \tag{55}$$

This circumstance clearly indicates that, unlike the $E$ - radial harmonics, the dispersion curve of each $H$ - radial harmonic contains only one branch. When condition (40) is satisfied, the cutoff frequencies of the $H$ - radial harmonics lie in the low-frequency region, but their values are higher than the low-frequency cutoff frequencies (44) corresponding to the $E$ - radial harmonics with the same ordinal numbers $n$. Near the cutoff frequency (52), from the dispersion equation (51) we obtain the dispersion law

$$\omega_{lf}^2 = \omega_{hn}^2 + \frac{k^2 c^2}{\varepsilon_f \mu_0}\left(1 + \frac{\mu_0 - 1}{\mu_0^3}\frac{\omega_{hn}^2}{\omega_h^2}\right),$$

In the vicinity of each cutoff frequency, the frequency increases with increasing longitudinal wave number.

Let us now consider the asymptotic behavior of the dispersion curves of the HF and LF waves in the short-wave limiting case $k \to \infty$. This limiting case corresponds to the quasi-longitudinal propagation of eigen electromagnetic waves in a coaxial ferrite waveguide. Let us first consider the behavior of the dispersion curves at $k \to \infty$ in the vicinity of the precession frequency $\omega \to \omega_h$. It is easy to verify that in this limiting case the condition $\rho_f \ll 1$ is satisfied and the expressions for the transverse wave numbers (29), (30) are valid. It is obvious that in this case $\lambda_1^2 \to -\infty$, and the value $\lambda_2^2$ remains finite so that $|\lambda_2^2/\lambda_1^2| \ll 1$. We also have the inequality $|\Gamma_1/\Gamma_2| = \rho_f \ll 1$. Taking into account the above, the dispersion equation (25) can be transformed to the form

$$\Delta_{11}(v_h\eta, v_h) - \frac{2}{\pi}\frac{\rho_f}{ka}\left[\frac{1}{\eta}\frac{J_1(v_h)}{J_1(v_h\eta)} + \frac{J_1(v_h\eta)}{J_1(v_h)}\right] = 0.$$

In the general case, the considered LF branch $\omega < \omega_h$ of the full dispersion curve looks as follows. This branch starts from a low cutoff frequency (40). At this frequency and in its vicinity, the eigen electromagnetic wave is an $E$ - wave. With increasing longitudinal wave number, the wave frequency increases. In this case, the contribution of the $H$ - component of the electromagnetic field is increased. At $k \to \infty$, the frequency asymptotically approaches the precession frequency from below, and the eigen electromagnetic wave becomes an $H$ - wave. For HF (optical) branches of natural electromagnetic waves, at $k \to \infty$, the frequency also increases unlimited $\omega \to \infty$. Since at $\mu_\perp \to 1$, and $|\mu_a| = (\mu_0 - 1)\omega_h / \omega \to 0$, then in this limiting case, the ferrite loses its magnetic properties. The system of eigen electromagnetic waves in the frequency range $\omega \gg \omega_h$ and their dispersion characteristics will be the same as in a coaxial waveguide completely filled with a dielectric with a permittivity of $\varepsilon_f$.

### 3. Dispersion of low-frequency TEM wave in layered coaxial ferrite line

To analyze numerical calculations of dispersion of ferrite coaxial lines with incomplete filling with ferrite, it is useful to have accurate analytical expressions for the phase velocity of the $TEM$ - wave (the initial slope of the dispersion curve) in the low-frequency limit. Such a calculation is presented below. The geometry of the line is described in Section 1. In the low-frequency range $\omega \ll \omega_h$, the gyrotropy of the ferrite can be neglected. The dispersion equation describing the $E$ - waves of the coaxial line under consideration has the form

$$\frac{\kappa_0}{\varepsilon_f} \frac{J_0(\kappa_0 r_f) N_0(\kappa_0 a) - J_0(\kappa_0 a) N_0(\kappa_0 r_f)}{J_1(\kappa_0 r_f) N_0(\kappa_0 a) - J_0(\kappa_0 a) N_1(\kappa_0 r_f)} = \frac{\lambda_d}{\varepsilon_d} \frac{J_0(\lambda_d r_f) N_0(\lambda_d b) - J_0(\lambda_d d) N_0(\lambda_d r_f)}{J_1(\lambda_d r_f) N_0(\lambda_d b) - J_0(\lambda_d b) N_1(\lambda_d r_f)}. \quad (56)$$

In the long-wave $kb \ll 1$ (low-frequency $\omega b / c \ll 1$) region, the dispersion of the $TEM$ - wave has the form of a straight line $\omega = k v_{ph}$, where $v_{ph}$ is the phase velocity of the $TEM$ - wave or the initial slope of the dispersion curve on the plane $(\omega, k)$. In this case, the transverse wave numbers for the ferrite and dielectric can be represented as

$$\kappa_0 = k q_f, \; q_f = \sqrt{\beta_{ph}^2 \mu_0 \varepsilon_f - 1}, \; \beta_{ph} = v_{ph}/c, \; \lambda_d = k q_d, \; q_d = \sqrt{\beta_{ph}^2 \varepsilon_d - 1}$$

and use the asymptotic representations of cylindrical functions for small values of the argument. As a result, the dispersion equation (56) is transformed to the form

$$q_f^2 \frac{\ln(r_f / a)}{\varepsilon_f} = -q_d^2 \frac{\ln(b / r_f)}{\varepsilon_d}.$$

From this equation we find expressions for the initial phase velocity $TEM$ - wave

$$v_{ph} = \frac{c}{\sqrt{\varepsilon_d \varepsilon_f}} \sqrt{\frac{\varepsilon_d \ln \frac{r_f}{a} + \varepsilon_f \ln \frac{b}{r_f}}{\mu_0 \ln \frac{r_f}{a} + \ln \frac{b}{r_f}}}. \quad (57)$$

Note that when the condition $\mu_0 \varepsilon_f > \varepsilon_d$ is met, the value of the phase velocity (57) is always within the limits

$$\frac{c}{\sqrt{\varepsilon_d}} > v_{ph} > \frac{c}{\sqrt{\mu_0 \varepsilon_f}}.$$

The upper border of this inequality corresponds to the phase velocity in a coaxial line completely filled with dielectric, and the lower limit of the phase velocity corresponds to a line completely filled with ferrite.

### Conclusion

By solving Maxwell's equations, an exact dispersion equation for electromagnetic waves propagating in a layered coaxial ferrite line is obtained. An analytical consideration is carried out for a simpler case of complete filling of the coaxial line with ferrite (a homogeneous ferrite line). In this case, the general dispersion equation splits into two independent equations. The first of them describes the dispersion of $TEM$ - electromagnetic waves, and the second one describes the dispersion of waveguide electromagnetic waves. Analysis of the dispersion equation for waves showed that the dispersion curve has two branches: LF and HF, which are separated by an opacity band $\mu_0\omega_h \geq \omega > \omega_h$. The LF branch begins with a linear section $\omega = kc/\sqrt{\varepsilon_f \mu_0}$. Then, with an increase of the wave number, the frequency increase slows down. At $k \to \infty$ the frequency of the LF *TEM* - wave asymptotically approaches to the precession frequency from below. The HF branch of the *TEM* - wave begins with a cutoff frequency $\mu_0\omega_h$. Then, with an increase in the wave number, the frequency increases and at $k \to \infty$ the dispersion curve of HF waves asymptotically approaches to straight line

$$\omega = \frac{c}{\sqrt{\varepsilon_f}}k + \frac{1}{2}(\mu_0 - 1)\omega_h.$$

In addition to *TEM* - waves, waveguide electromagnetic waves can propagate in a ferrite coaxial line. These waves, unlike *TEM* - waves, have cutoff frequencies. In the region of long waves, where the gyrotropy of ferrite can be neglected, the total electromagnetic field is divided into $E$ and $H$ - waves. Each $E$ - radial harmonic has two cutoff frequencies: low-frequency and high-frequency. These cutoff frequencies correspond to the LF $\omega < \omega_h$ and HF $\omega > \omega_h\mu_0$ branches of the dispersion curve. With increasing wave number, the frequencies of the LF and HF radial harmonics increase. At $k \to \infty$, the dispersion curve of each LF $E$ - radial harmonic, as in the case of *TEM* - waves, asymptotically approaches the precession frequency from below. The frequency of each HF $E$ - radial harmonic also increases with increasing longitudinal wave number and at $k \to \infty$ asymptotically approaches the straight line $\omega = kc/\sqrt{\varepsilon_f}$, since the ferrite loses its magnetic properties in the frequency range $\omega \gg \omega_h$.

Radial $H$ - harmonics have only one cutoff frequency. This circumstance clearly indicates that, unlike $E$ - radial harmonics, the dispersion curve of each $H$ - radial harmonic contains only one branch. With the growth of the longitudinal wave number, the frequency monotonically increases, intersects the resonance line $\omega = \omega_h$ and at asymptotically approaches the same line $\omega = kc/\sqrt{\varepsilon_f}$, as in the case of $E$ - waves.

The expression is obtained for the phase velocity of the *TEM* - wave in the LF range for a layered ferrite - dielectric coaxial line. Thus, the boundaries are determined, within which the phase velocity of the *TEM* - wave changes for a specific geometry of the ferrite line.